\documentclass{kapproc} 

\usepackage{procps} 

\usepackage[dvips]{graphicx}

\upperandlowercase

 \let\footnote\savefootnote

\let\footnotetext\savefootnotetext

\let\footnoterule\savefootnoterule

\kluwerbib 

\begin{document}

\articletitle[The Galaxy Structure-Redshift Relationship]
{The Galaxy Structure-Redshift Relationship}

\author{Christopher J. Conselice\altaffilmark{1}}
 
\affil{\altaffilmark{1}California Institute of Technology, Pasadena, CA USA}  

\begin{abstract}

There exists a gradual, but persistent, evolutionary effect in the galaxy
 population such that galaxy structure and morphology change with redshift.
This galaxy structure-redshift relationship is such that an increasingly large
fraction of all bright and
massive galaxies at redshifts $2 < z < 3$ are morphologically peculiar at  
wavelengths from rest-frame ultraviolet to rest-frame optical. There are 
however examples of morphologically selected spirals and ellipticals at
all redshifts up to $z \sim 3$. At lower
redshift, the bright galaxy population smoothly transforms into normal
ellipticals and spirals.  The rate of this transformation 
strongly depends on redshift, with the swiftest evolution occurring between 
$1 < z < 2$. This review characterizes the galaxy structure-redshift 
relationship, discusses its various physical causes, and how these
are revealing the mechanisms responsible for galaxy formation.

\end{abstract}


\section{Introduction}

The structures and morphologies of galaxies change with time.
Determining the history and cause of this galaxy structure-redshift
relationship, including the origin of 
modern galaxy morphologies (i.e., ellipticals, disks) is perhaps
the missing, and until now overlooked, link in understanding galaxy 
formation. While
we currently have a good understanding of global 
galaxy formation and evolution, such as the star formation and
mass assembly history (e.g., Madau et al. 1996; Dickinson et al. 2003), we are 
only beginning to understand {\it how} galaxy formation occurs as opposed to
simply when. During the last few years it has become clear with the advent
of wide-field imaging surveys from the ground, and from
space using the Hubble Space Telescope, that galaxy structure evolves 
(e.g., Driver et al. 1995; Glazebrook et al. 1995; Abraham et al. 1996;
Brichmann \& Ellis 2000;
Conselice et al. 2004b).  There is a clear galaxy structure 
(or morphology)-redshift relationship
such that galaxies in the more distant universe are peculiar while those
in the local universe are more regular or normal\footnote{Throughout this
review, I refer to 'regular' or 'normal' galaxies to denote
systems that are on the present day Hubble sequence; namely 
ellipticals and spirals.  Peculiar galaxies and/or mergers
are not normal galaxies according to this criteria. This differs from
the usual meaning of normal which refers to galaxies without the presence
of an active galactic nuclei (AGN).} 
(van den Bergh et al. 2001).  Determining the physics behind the 
morphology-redshift relationship is critical for any ultimate understanding
of galaxies and the physical causes of galaxy structure and its evolution.

The morphology-redshift relationship can furthermore potentially be used as 
a key test
of galaxy formation models.   Theories of galaxy formation
can be divided into two main ideas - the monolithic collapse of material
early in the universe to form stars and galaxies within a very short
time (e.g.,
Larson 1975; Tinsley \& Gunn 1976) and the hierarchical 
formation scenario (e.g., White \& Rees 1978; Blumenthal et al. 1984; White
\& Frenk 1992; Cole et al. 2000). Observationally,
we know that galaxies do not
appear to form rapidly in the early universe, but have  an extended star
formation history that does not decline significantly until
the universe is about half its current age. Likewise, about half of
all stellar mass in the universe formed between $z \sim 1$ (8 Gyrs ago) and
today (Dickinson et al. 2003).   The fact
that star formation occurs over time, and not quickly at very high redshift,
largely rules out rapid collapses as the primary method for forming all
galaxies.   High redshift galaxies also tend to be small with likely small
stellar masses (Papovich et al. 2001; Ferguson et al. 2004).  
Therefore a large fraction of all galaxies must have formed gradually throughout time.
Understanding this process, that is what is causing mass to build up
in galaxies, requires studying their internal properties.

There are several ways to measure the physical processes responsible for
forming galaxies which can potentially explain the observed morphology-redshift
relationship (\S 3).  One method, and by far the most common, is to
study global galaxy properties, such as the evolution of stellar mass and
star formation, and to compare these with predictive models (e.g., Somerville
et al. 2001).  Other methods, which are now just being explored, involve 
probing the internal features of
high-z galaxies either through spectroscopy or high resolution
imaging.  While integral field spectroscopy for high-z galaxies is still
in its infancy, understanding the internal structural features of high
redshift galaxies in now in a golden age, utilizing new techniques (e.g., 
Conselice et al. 2000a; Peng et al. 2002) with high resolution Hubble
Space Telescope imaging (e.g., Giavalisco et al. 2004; Rix et al. 2004).

The idea that the structures of galaxies hold clues towards
understanding their current and past formation histories is a new,
and perhaps still controversial, idea.  There is however increasing amounts of
evidence that suggests galaxy structure reveals fundamental past
and present properties of galaxies (see Conselice 2003 and references 
within and \S 2).   Utilizing these tools, we can begin to determine 
the origin of 
the galaxy morphology-redshift relationship.  Understanding this relationship
will in turn help us determine physical formation mechanisms, and the 
history of 
galaxy assembly.   Furthermore, it is
now possible to compare observations of the galaxy morphology-redshift
relationship with theoretical models based on cosmological and dark matter 
ideas, connecting the universe as a whole to its constitute galaxies.    
I argue in this review that the galaxy morphology-redshift
relationship is a pillar for understanding galaxy formation. It may also
hold clues for understanding the relationship between the
baryonic content of galaxies and their dark matter halos, the evolution of
galaxies and their black holes, and the relationship between 
cosmological parameters and the evolution of galaxy structure.
In summary, I will address in detail the following issues: 

\noindent (i) What is the galaxy structure-redshift relationship and how does it evolve?
\noindent (ii) What is the physical causes for the formation and evolution of
     the galaxy structure-redshift relationship? \\
\noindent (iii) What does the evolution of the galaxy-structure redshift 
relationship tell us about galaxy formation?

I do not address some morphological evolution problems, such
as the formation and evolution of bars, rings or other internal
structures, as these are dealt with in other contributions (e.g., Jogee,
Sheth). 
Throughout I will assume a cosmology with H$_{0} = 70$ km s$^{-1}$
Mpc$^{-1}$ and relative densities of $\Omega_{\Lambda} = 0.7$ and
$\Omega_{\rm m} = 0.3$.  

\section{The Physical Basis of Galaxy Structure}

Before describing the galaxy morphology-redshift relationship, and 
what it implies for understanding galaxy formation,
I will review our current understanding of how galaxy structure
correlates with physical properties of galaxies. It has been known
for decades that, broadly speaking, galaxy morphology correlates with 
galaxy properties such as luminosities, sizes, gas content, 
colors, environment, masses, and mass to light ratios (e.g.,
Roberts \& Haynes 1994).  Generally, early-type galaxies (ellipticals)
are larger, more massive, contain older stellar populations, and are
found in denser areas than spirals.  Later type galaxies, such as spirals, 
are bluer,
contain younger stellar populations, more gas, and are less massive overall
than the ellipticals.  The detailed correlation between these physical 
properties and
Hubble/de Vaucouleur classifications is however not strong.  While
in the mean properties change with Hubble type, there is
significant overlap in any given property across the Hubble sequence\footnote{Room prohibits a detailed discussion of all the problems with  
Hubble classifications.  
The fact that physical properties only correlate in the mean for a given
Hubble type is only one of many issues.  For a detailed discussion of this
see Appendix A from Conselice (2003).  It is still useful to separate in
the broadest sense, ellipticals from spirals, as I do here, as these are 
fundamentally different galaxy types.}.

\begin{table}[ht]
\caption[The co-moving densities of bright (M$_{\rm B} < -20) galaxies as a 
function of morphological type]
{The co-moving densities of bright (M$_{\rm B} < -20$) galaxies as a function of morphological type in units of log (Gpc$^{-3}$)$^{a}$}
\begin{tabular*}{\textwidth}{@{\extracolsep{\fill}}lccc}
\sphline
Redshift & Ellipticals & Spirals & Peculiars \cr
\sphline
       0.0   &    6.41$\pm$0.01 & 6.98$\pm$0.01 & 5.34$\pm$0.03 \cr
       0.5   &    6.3$\pm$0.2 & 6.6$\pm$0.1  & 5.7$\pm$0.2\cr
       0.8   &    6.3$\pm$0.2 & 6.2$\pm$0.2  & 5.4$\pm$0.3\cr
       1.0   &    6.4$\pm$0.2 & 6.5$\pm$0.2  & 6.1$\pm$0.2\cr
       1.2   &    6.2$\pm$0.2 & 6.1$\pm$0.2  & 5.7$\pm$0.3\cr
       1.3   &    6.0$\pm$0.2 & 5.9$\pm$0.2  & 6.3$\pm$0.2\cr
       1.5   &    5.7$\pm$0.2 & ...      & 5.4$\pm$0.3\cr
       1.6   &        ...     & 5.4$\pm$0.3  & 5.7$\pm$0.3\cr
       1.7   &    5.4$\pm$0.3 & 5.7$\pm$0.2  & 6.1$\pm$0.2\cr
       1.9   &    5.7$\pm$0.2 & 5.4$\pm$0.4  & 5.7$\pm$0.3\cr
       2.0   &    5.9$\pm$0.2 & 6.1$\pm$0.2  & 6.4$\pm$0.1\cr
\sphline
\end{tabular*}
\begin{tablenotes}
$^a$Galaxies at $z = 0$ are taken from the ``Third Reference Catalogue of
Bright Galaxies'' (de Vaucouleurs et al. 1991).  The other galaxy densities
at $z > 0$ are an average between the densities found for each type in the 
Hubble Deep Fields North and South.
\end{tablenotes}
\end{table}

Some quantitative measurements of galaxy structure, on the other hand, 
correlate strongly with physical properties, including: the current
star formation rate, the stellar mass, galaxy radius, central black hole mass, 
and merging properties.  It is impossible to describe all but a few of these
correlations here.  The first of these discovered 
is that the light concentration of an evolved stellar population correlates 
with its luminosity, stellar mass, and scale (e.g., Caon, Capaccioli \& 
D'Onofrio 1993; Graham et al. 1996; Bershady et al. 2000; Conselice 2003).  This
was first noticed by the failure of the de Vaucouleurs r$^{1/4}$ surface
brightness profile to fit the surface brightness distributions of elliptical
galaxies brighter than or fainter than M$_{\rm B} \sim -20$ (e.g., Schombert
1986).  It was realized later that the shape of the surface brightness
profile for early types correlates strongly with its absolute
blue magnitude (Binggeli \& Cameron 1991; Caon et al. 1993).  It was
later shown that the general Sersic profile, with its concentration
parameter n, gives a much better
fit than the de Vaucouluer's profile for all early types (Graham et al. 1996).
The central light concentration, as measured through the Sersic n index,
or a non-parametric concentration index ($C$) (Bershady et al. 2000),
also correlates
with the mass of central black holes (Graham et al. 2001).

While the concentration of a galaxy's light profile correlates with its
stellar mass, absolute magnitude, and size, in a sense revealing
the past formation history of a galaxy, there are many indicators in
the structures of galaxies for ongoing galaxy formation.
For example, recently it has been shown that the clumpiness of a galaxy's light
distribution correlates with the amount and the location of star formation 
(e.g., Takamiya 1999; Conselice 2003).   The clumpiness 
is measured by quantifying the fraction of a galaxy's light in the rest-frame
B-band in high spatial frequency structures.  The ratio between
the amount of light in these high spatial frequency structures and the total
light gives a measure of the clumpiness, or star formation.  This trend can be 
demonstrated by the strong
correlation between the clumpiness index $S$ (Conselice 2003) and 
H$\alpha$ equivalent widths and colors of star forming galaxies. 
There is also a strong
relationship between the dynamical state of a galaxy and the presence of a
merger.  Generally, merging galaxies are asymmetric, while non-mergers are
not (Conselice et al. 2000a,b;
Conselice 2003). This has been shown in numerous ways, including
empirical methods (Conselice 2003) and the correlation of internal 
HI dynamics and asymmetries of stellar distributions (Conselice et al. 2000b). 

\begin{figure}[ht]
\centerline{\includegraphics[width=4.5in]{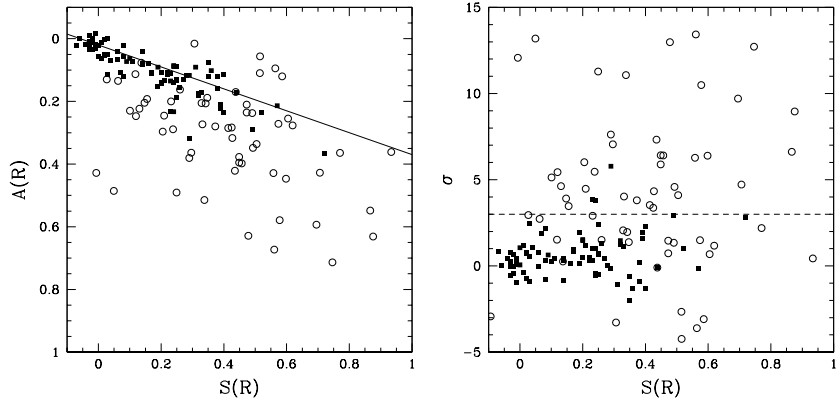}}
\caption{\protect\inx{} The relationship between the asymmetry index
($A$) and the clumpiness index ($S$). Higher $A$ and $S$ values indicate
galaxies that are more asymmetric and have a higher fraction of clumpy light,
respectively.  For normal galaxies (black squares,
left panel) there is a strong correlation between $A$ and $S$ such that
$A = (0.35\pm0.03) \times S + (0.02\pm0.01)$.  The galaxies which
deviate from this relationship are the ongoing major mergers, shown in 
the left panel as open circles.  The right panel shows the deviation from
the $A-S$ relationship in units of the scatter of the asymmetry values
of the normal galaxies ($\sigma$).
Generally, only the mergers deviate from this relationship by more
than 3$\sigma$. For a physical reasoning behind these correlations see the
text and Conselice (2003).}
\end{figure}

As shown in Conselice et al. (2000a) asymmetric light distributions can
also be caused by star formation.  However, by decomposing light and
kinematic structures in galaxies, it is possible to show that primary 
asymmetries are not the result of star formation, which forms in clumps 
(e.g., Elmegreen 2002), but from large scale lopsidedness on the order of
the size of 
the galaxy itself (Andersen et al. 2001).  Likewise, there is a strong 
correlation
between the asymmetry parameter and the clumpiness parameter for normal
star forming galaxies (Conselice 2003; Figure~1).  Galaxies with high clumpiness
values ($S$), which correlates with high amounts of star formation, 
have correspondingly higher
asymmetry values.  However, this correlation breaks down for systems
involved in major mergers, such as nearby ultraluminous infrared galaxies
(Figure~1). The nature of this deviation is such that a galaxy undergoing
a merger has too high an asymmetry for its clumpiness, 
demonstrating that large asymmetries are produced in large scale features
and not in clumpy, star formation like regions (Conselice 2003; Mobasher et al.
2004; Figure~1)\footnote{The
concentration index ($C$), asymmetry index ($A$), and 
clumpiness index ($S$) form the CAS morphology system described in detail
in Conselice (2003).  With these three parameters the major classes of nearby
galaxies can be distinguished.  Although this idea is not explicitly
described in this review, it is used in papers described here, 
such as Conselice et al. (2004b) and Mobasher et al. (2004) to determine 
galaxy types.}.

\section{The Galaxy Morphology-Redshift Relationship}

\subsection{Summary}

The summary figure for understanding the galaxy structure-redshift relationship
is Figure~2.  This relationship can be summarized simply 
as: {\em At higher redshifts 
(early times) the fraction of bright galaxies that are peculiar in structure
and morphology increases 
gradually at the expense of both spirals and ellipticals.}

\begin{figure}[ht]
\centerline{\includegraphics[width=4.5in]{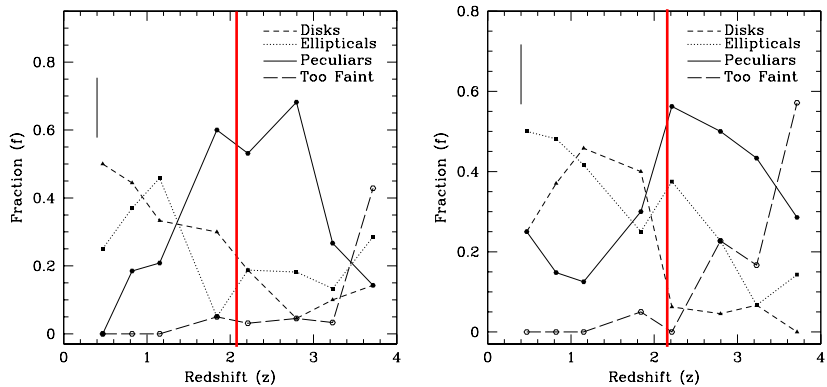}}
\caption{\protect\inx{\bf The galaxy structure-redshift relationship.} This
figure shows the evolution in relative fractions of different galaxy types
as a function of redshift for classifications in the I$_{814}$ (left panel)
and H$_{160}$ (right panel) band images of the HDF-N.  The short vertical 
solid line on each plots gives the average
error for these fractions. The long vertical line is the redshift limit
for detecting spirals and ellipticals with M$_{\rm B} < -20$.  
Types plotted are: disks (short dashed), ellipticals (dotted),
peculiars (solid) and galaxies which are too faint for a classification 
(long dashed). }
\end{figure}

The final state of galaxy evolution surrounds us, and the modern universe is
dominated by galaxies that can be classified on the Hubble sequence. A large
fraction of all modern massive and bright galaxies are either ellipticals
or spirals; only roughly 1-2\% of all bright galaxies with M$_{\rm B} < -20$
can be classified
as peculiars (e.g., Marzke et al. 1998) (see Table~1).  
This changes gradually with redshift up to
$z \sim 1$ and then more rapidly between $1 < z < 2$; 
at $z \sim 1$ most galaxies
have relaxed morphologies while at $z \sim 2$ most galaxies are peculiar
(Conselice et al. 2004b; Table 1).

\subsection{Galaxy Structures at Low Redshift $z < 1$}

At redshifts $z < 1$ most of the bright (M$_{\rm B} < -20$) and
massive galaxies (M$_{*} > 10^{10}$) are normal galaxies, that is
ellipticals and spirals (Table~1; Figure~2).  This relative fraction remains
largely similar out to $z \sim 1$, with some important exceptions.  In
general, the co-moving density of elliptical and disk galaxies
remains constant, to within a factor of 2, out to $z \sim 1$ with
a slight decline (Figure~3;
Brinchmann \& Ellis 2000; Conselice et al. 2004b).  

There is a more pronounced change in other features of normal
galaxies from $z \sim 1$ to $z \sim 0$.  These properties include 
co-moving B-band luminosity densities
($\rho_{\rm L_{B}}$), and stellar mass densities ($\rho_{*}$).  
While the number density evolution of Hubble types 
is the physical manifestation of the galaxy structure-redshift
relationship, the evolution of other properties can reveal important
clues for how this relationship is put into place, and why it might
be evolving.  
The rest-frame B-band luminosity luminosity density evolution for galaxies of 
known morphology
is shown in Figure~3.  There is a clear decline with cosmic time
 in luminosity densities
at $z < 1$ for all galaxies,
including ellipticals and spirals.  This peak in the B-band luminosity
density at $z \sim 1$ is produced in normal galaxies, and must be due to 
recent star formation, as the stellar mass
density for normal galaxies grows with time (Figure~3).   
The stellar mass density for ellipticals is half of its modern value at
$z \sim 1$ in the Hubble Deep Field North (HDF-N).  There is perhaps
an over density of ellipticals at $z \sim 1$ in the HDF-N, and cosmic
variance is an issue. Although a lower density of early types would only
enhance the evolution in stellar mass for these systems.  This effect is
also seen in studies considering galaxies on the `red sequence', defined
by the tight correlation between magnitude and color for early types.   
The stellar mass in red sequence galaxies increases by a factor
of two from $z = 1$ to $z = 0$ (Bell et al. 2004), exactly the increase
found when considering morphologically selected early types. Because
of the large amount of co-moving luminosity in normal galaxies at $z\sim 1$,
star formation
must be occurring in early type galaxies during this time (see also
Stanford et al. 2004). Luminosity and
stellar mass functions suggest that this evolution is occurring in lower 
mass and lower luminosity systems (Conselice et al. 2004b), while
the higher mass systems are perhaps largely formed by $z \sim 1$, or even
earlier (e.g., Glazebrook et al. 2004).

\begin{figure}[ht]
\centerline{\includegraphics[width=4.5in]{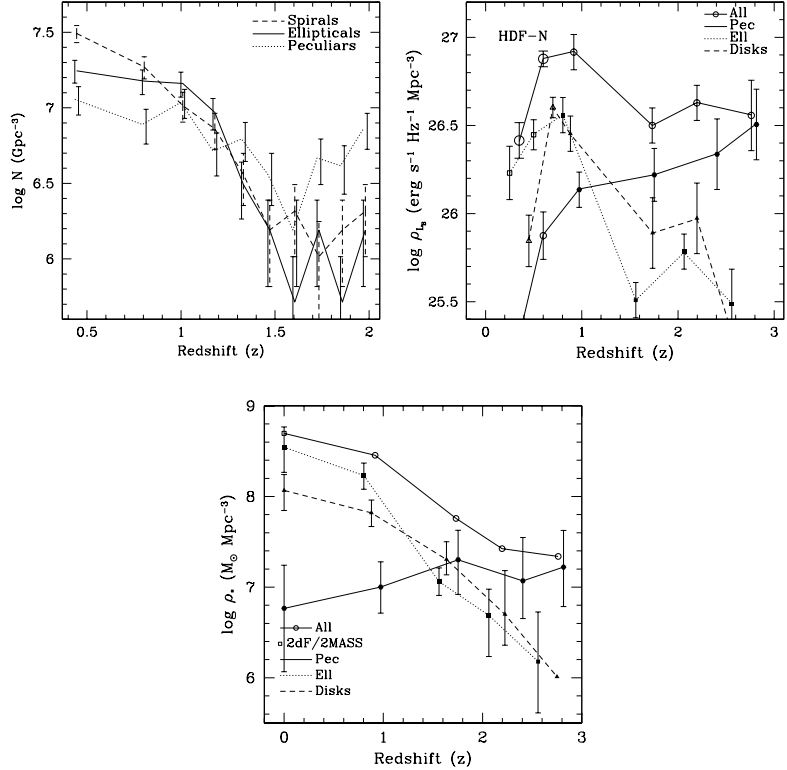}}
\caption{\protect\inx{} The relative co-moving number (N), 
rest-frame B-band luminosity ($\rho_{\rm L_{B}}$), and
stellar mass density ($\rho_{*}$) of galaxies as a function of 
redshift from deep NICMOS imaging
of the Hubble Deep Field North (Conselice et al. 2004b).  Points
at redshifts $z < 0.5$ are taken from Brinchmann \& Ellis (2000), 
Fukugita et al. (1998) and the 2dF/2MASS surveys (Cole et al. 2001).}
\end{figure}

\subsection{Galaxy Structures at Medium Redshift $1 < z < 2$}

Morphological counts of galaxies from early Hubble Space Telescope
imaging found a large increase in the number of 
peculiar/irregular galaxies at fainter magnitudes (e.g., Driver et al. 1995;
Glazebrook et al. 1995).    It was however unknown during
these early Hubble observations what the redshifts, and therefore the 
characteristics, of these peculiar 
galaxies were.  When redshifts for these faint galaxies became available,
it was argued that Hubble types appeared in abundance by $z \sim 1$,
and evolve only slightly down to lower redshifts (van den Bergh et al. 2001;
Kajisawa \& Yamada 2001).

Ultimately what is desired is a determination of Hubble types as a
function of redshift for galaxies of different luminosities and
stellar masses.  This was performed for bright galaxies in the
Hubble Deep Field North and South by Conselice et al. (2004b).  The results
of this are shown in Figure~3 for galaxies brighter than I = 27.    
As described in \S 3.1 there is a rapid decline with increasing redshift
in the number of normal
galaxies between $z \sim 1$ and $z \sim 1.5$, such that the co-moving
density increases by 8.3 $\times 10^3$ Gpc$^{-3}$ Gyr$^{-1}$ for
ellipticals and 5.7 $\times 10^3$ Gpc$^{-3}$ Gyr$^{-1}$ for spirals during
this 1.6 Gyr period, roughly a factor of 10 increase in number densities.  
As discussed briefly in \S 3.5 this
change in morphology is not caused by so-called morphological k-corrections
in which galaxies appear different at different wavelengths.  It can however
be partially produced by selection effects, although a strong drop is
also found when considering galaxies at a fixed absolute magnitude (Figure~2). 
Both spirals and ellipticals with M$_{\rm B} < -20$ should be found
in the Hubble Deep Fields up to $z \sim 2$, and at even higher redshifts
if passive evolution is considered (e.g., Conselice et al. 2004b).

The redshift range $1 < z < 2$ is obviously critical for understanding
the final onset and production of the Hubble sequence and the origin of
the galaxy structure-redshift relationship.  It is also the epoch (during a 
short 2.5 Gyrs!) where the star formation rate, AGN activity and stellar mass
assembly is at its highest.  Understanding how the galaxy structure-redshift
relationship evolves during this epoch is critical for understanding
the causes behind
galaxy formation.  It is therefore worth spending some time discussing
what is found morphologically and structurally in the galaxy
population between $1 < z < 2$.   

Figure~4 shows Advanced
Camera for Surveys (ACS) images of the brightest galaxies in the rest-frame
optical found within the Chandra Deep Field South Great Observatory Origins
Survey (GOODS) imaging.  Clearly, there is a rich morphological mix
at this redshift, with many galaxies appearing similar to modern ellipticals
and spirals, but with important structural differences that make them
fundamentally different from modern normal galaxies. Another
way to investigate this population is to study systems that have spectral
energy distributions that likely place them at $1 < z < 2$, such as
the extremely red objects, discussed in \S 4.1.  The images of these
galaxies reveal that some are almost normal, with outer shell like
features and what
appear to be large star forming complexes. Understanding the physical causes
behind these features will help reveal the formation mechanisms of galaxies.

\begin{figure}[ht]
\centerline{\includegraphics[width=5in]{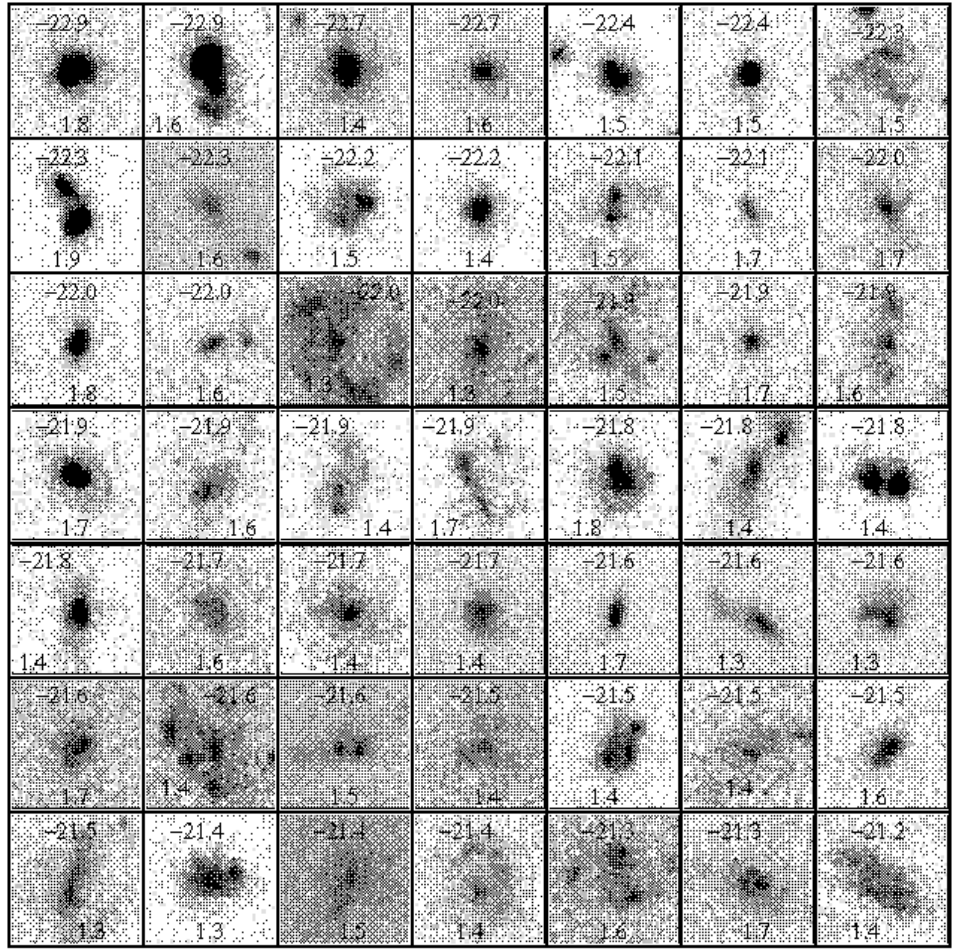}}
\caption{\protect\inx{}  The brightest galaxies in ACS 
GOODS images whose photometric redshifts place them at $1 < z < 2$.  These
are ordered from brightest to faintest down to M$_{\rm B}$ = $-21$.  The
upper number is the M$_{\rm B}$ of each galaxy and the lower number is its
redshift.  There is  a large
diversity of properties, from systems that appear very peculiar
to those that look similar to normal galaxies.   Scale of these images is 
$\sim 2''$  on each side, corresponding to $\sim$17 kpc at these redshifts. \bf {See Figure4.gif for a high resolution version.}}
\end{figure}

\subsection{Galaxy Structures at High Redshift $z > 2$}

The structures and morphologies of galaxies at $z > 2$ are just now
being studied in detail.  Early work in this area suggested that
galaxies selected by the Lyman-break technique have compact structures 
with outer light envelopes (Giavalisco et al. 1996).
These compact structures have half-light
radii a few kpc in size, similar to the bulges of modern spirals or moderate
luminosity spheroids.  Many of these compact galaxies have steep light
profiles and asymmetrically distributed outer nebulosity. 
The depth of this early imaging was
quickly superseded by the Hubble Deep Field North which
showed a rich diversity of galaxy structures (Figure~5; 
Ferguson, Dickinson \& Williams
2000).  The morphologies of these galaxies is still however a largely 
unexplored area of parameter space.

\begin{figure}[ht]
\centerline{\includegraphics[width=4in]{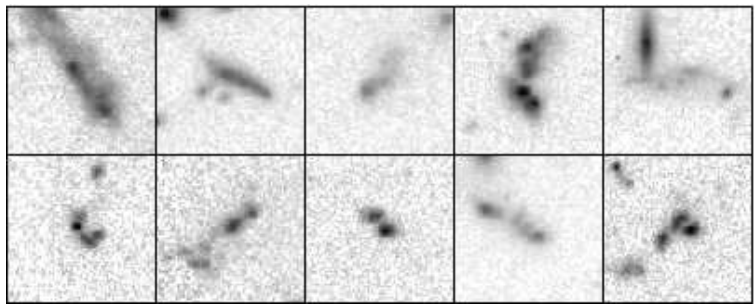}}
\caption{\protect\inx{} The morphologies of bright galaxies, mostly at $z > 2$,
in the Hubble Deep Field North showing the peculiar and non-compact structures
of these galaxies, many with several bright central regions or knots.}
\end{figure}

One of the reasons the morphologies of $z > 2$ galaxies have not been
studied in detail is that describing their structures  is not a trivial
problem, as very few galaxies at high redshift can be identified as
objects that would fit on the Hubble Sequence (e.g., Giavalisco et al. 1996;
Conselice et al. 2003a; Lotz et al. 2003).  One way to approach this problem 
is to use a purely descriptive approach described above, while
another is to use quantitative techniques to characterize these
structures.  The has been done in Conselice et al. (2003a) and 
Conselice et al. (2004b) for galaxies at $z > 2$.  The CAS systems
shows that at the highest redshifts there is a real dichotomy in the
galaxy population such that the most luminous and most massive galaxies
are consistent with undergoing a major merger based on CAS indices,
particularly the asymmetry index (Conselice et al. 2003a).  What is found
is that half of all bright, M$_{\rm B} < -21$ or massive M$_{*} > 10^{11}$
M$_{\odot}$ galaxies are actively undergoing a major merger.  The fainter and
lower mass galaxies have peak merger fractions at $z \sim 2.5$ that
are only 20\%, or lower.  The relative fraction of mergers declines at
lower redshifts very quickly for these massive and luminous galaxies
as power law $\alpha$ (1+z)$^{3-5}$ (Conselice et al. 2003).
A comparison to hierarchical assembly models of galaxies is shown in
Figure~6 using  GALFORM simulations (e.g., Benson et al. 2002).   These 
models over-predict the number of major mergers occurring for
the brightest galaxies at $z < 1$, consistent with the fact that there
are too many bright K-band selected galaxies at $z \sim 1 - 1.5$
than predicted in Cold Dark Matter based models (Somerville
et al. 2004; see also \S 4.3).

\subsection{Morphological K-corrections}

A primary problem in understanding the galaxy structure-redshift relationship
is constraining the effects of the morphological k-correction, whereby galaxy
structure changes as a function of wavelength.  The main problem is that most
deep high resolution imaging is done in the optical, probing
up to $\lambda = 1 \mu$m, allowing for a sampling of the rest-frame
optical, $< 4000$ \AA, only up to $z \sim 1.5$.  At redshifts higher than this
we begin to sample rest-frame ultraviolet light from galaxies.  
For galaxies at $z \sim 3$, for example, the z-band filter (F850L) (the
reddest GOODS and ACS Hubble Ultra Deep Field filter) samples
$\sim$ 2500 \AA, the near ultraviolet, where only young stars with ages 
$< 100$ Myrs are sampled.  

\begin{figure}[ht]
\centerline{\includegraphics[width=4in]{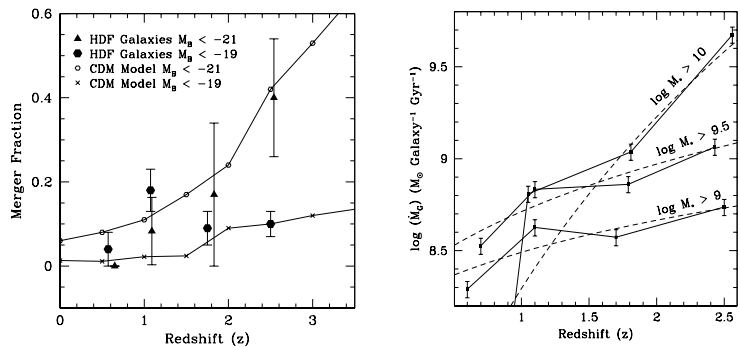}}
\caption{\protect\inx{} Left panel: major merger fractions to 
$z \sim 3$ at
magnitude limits M$_{\rm B} = -21$ and $-19$. Semi-analytical model
predictions are also shown.  Right panel: Stellar mass accretion
history from major mergers as a function of initial mass 
(see Conselice et al. 2003a). }
\end{figure}

This is potentially a problem because both qualitatively, and quantitatively, 
galaxies have very different
structures in their rest-frame optical and UV light (e.g., Bohlin et al.
1991; Kuchinski et al. 2001; Windhorst et al. 2002; Papovich et al. 2003).
The largest differences are found for galaxies that are composed of
old and young stellar components which are not spatially mixed, such as 
early type
spirals or Hubble classifications Sa and Sb (e.g., Windhorst et al. 2002).
The one type of galaxy that looks nearly identical at UV and optical
wavelengths are starbursting galaxies, or galaxies whose structures
are dominated by star formation (e.g., Conselice et al. 2000c; Windhorst
et al. 2002).

Similarly, when examining galaxies at different redshifts, there are
different types of morphological k-corrections, depending on the redshift
and galaxy type.  At $z < 1.5$ morphological k-corrections are not
an issue as we are able to sample rest-frame optical light from galaxies.  
At higher
redshifts the NICMOS camera on HST allows us to determine the rest-frame
optical structures and morphologies, although only limited field
coverage exists (e.g., Dickinson et al. 2000).  The result of this imaging
is that galaxies that look irregular and distorted in the rest-frame UV
also appear distorted in the rest-frame optical (Teplitz et al. 1998; 
Thompson et al. 1999), with some possible
and important exceptions (e.g., Giavalisco 2002; Labbe et al. 2003; 
Conselice et al. 2004a,b).   While these are rare systems, 
they do exist, and examples of morphologically selected ellipticals are found
out to $z > 2$.  These normal galaxies will likely become more common 
as we probe deeper in the infrared with high resolution wide field
imaging.

\section{The Nature of the Peculiar High-z Galaxy Population}

\subsection{Extremely Red Objects}

Extremely red objects (EROs), defined by an optical-infrared color
limit, typically using
the criteria (R-K) $> 5-6$, or (I-K) $> 4-5$ (e.g., Daddi et al. 2000),
were traditionally thought to be early type or dusty galaxies at $z > 0.8$.  
Extremely red galaxies
are red because of a large 4000 \AA\,  break produced
by aged, or dusty, stellar populations.  EROs are now also found in the near
infrared with a (J-K) limit that locates objects at redshifts $z > 2$.  The
ERO population is therefore a good one for determining the basic properties
of evolved galaxies at high redshifts.

The morphologies of EROs are mixed, with a strong redshift dependence 
(Figure~7).  Systems
at $z < 1.2$ are typically early or late-type galaxies, while those at 
$z > 1.2$
are more irregular or peculiar (Moustakas et al. 2004).  The
spectra of EROs are also mixed, with about half showing signs of evolved
stellar populations, with the other half showing emission lines (Cimatti
et al. 2002).  Morphological studies of EROs demonstrate that a large fraction,
perhaps the majority of the K $< 20$ objects, are disks (Yan
\& Thompson 2003).  It thus appears likely that the ERO definition, far
from finding only specific galaxy populations, is sampling all morphological
types.  This is a good sample for studying  galaxies
evolving onto the Hubble sequence since Hubble types are the most evolved
galaxies at $z \sim 0$ and are likely also the most evolved at $z \sim 1$. 
This has been done by e.g., 
Moustakas et al. (2003) who studied the redshift distributions of
EROs in the GOODS-South field as a function of morphology.  Moustakas
et al.  found that the lower redshift EROs are dominated by regular
galaxies, while higher redshift samples at $z > 1.5$ are 
dominated
by galaxies that cannot be place on the Hubble sequence (Figure~7).  

\begin{figure}[ht]
\centerline{\includegraphics[width=3in]{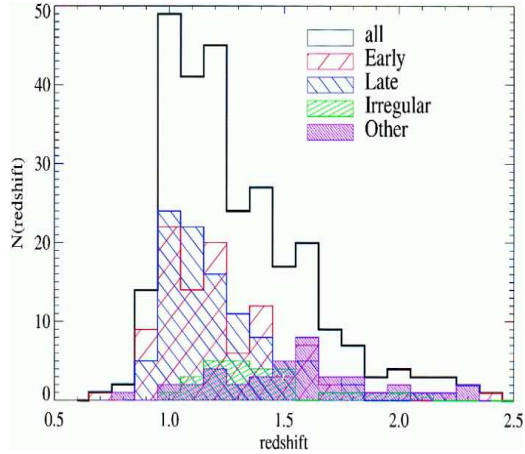}}
\caption{\protect\inx{} The redshift distributions for extremely
red galaxies from the GOODS-South field (Moustakas et al. 2004) showing
how the morphological distribution for this popualtion shifts to the 'other'
or peculiar types at higher redshifts.   }
\end{figure}

Recently, there have been claims of a new population of 
red galaxies at $z > 2$.  These
are similar to the lower redshift ERO population in that they are identified
by a color cut that isolates galaxies based on the Balmer break, but uses
the color defined by two infrared filters, normally the J and K bands (e.g.,
(J-K) $> 2.3$).
These systems are typically at $z > 2$ and may be quite distinct from the 
Lyman-break galaxy population (Franx et al. 2003; van Dokkum et al. 2004).
The morphologies of these systems have not been studied in detail, partially
because there are so few examples, yet bright
K-band selected galaxies at $z > 2$ and with K $< 20$ often have irregular 
UV morphologies
(Labbe et al. 2003; Daddi et al. 2004) indicating star formation.    There
are however some hints for spiral structures and more regular compact
near infrared morphologies in some of these systems (Labbe et al. 2003; 
Daddi et al. 2004).   These infrared EROs have 
large stellar masses and are generally consistent, based on clustering 
analyses and stellar population arguments, with the most evolved systems at 
high redshift.  These galaxies are therefore the best candidates for 
being the progenitors of evolved galaxies found in dense regions today.

\subsection{Luminous Diffuse Objects and Chain Galaxies}

In Conselice et al. (2004) a new galaxy type, found abundantly between
$1 < z < 2$, is described and characterized.  These galaxies, which
have no local counterparts, were discovered based on their low light
concentrations and high luminosities, and are called Luminous Diffuse
Objects (LDOs).  These objects are fairly common with surface
densities 1.8 arcmin$^{-2}$, and co-moving number densities of $5 \times
10^{5}$ Gpc$^{-3}$ within the GOODS South 
field  (Conselice et al. 2004).  These objects were independently discovered by
Elmegreen et al. (2004a) and are likely face on counterparts of the
`chain galaxies' discussed in Cowie et al. (1995) (Elmegreen et al.
2004b).

 Elmegreen
et al. (2004b) compared the colors of the star forming knots in their sample
of LDOs with
the knots found in chain galaxies, finding a very similar color distribution.
This suggests that chain galaxies are the edge-on versions of LDOs.  
Both chain galaxies and LDOs are known to be large complexes of star forming 
regions (Cowie et al. 1995; Conselice et al. 2004a).    Formation scenarios for
these systems are discussed in Elmegreen et al. (2004b), and are consistent
with large amounts of star formation occurring after gas in an initial
disk fragments and produce several large clumps.  In models, these
clumps are predicted to form through energy dissipation and later
merge together to form bulges (Immeli et al. 2004).  The fact that
there are no obvious bulge components in LDOs is a clue that bulge
formation may occur {\em after} disk formation, not before, as is generally
assumed in hierarchical models.

LDOs are likely in a phase where a large fraction of their stellar
mass is being assembled.  The star formation rate in LDOs is
on average 4 M$_{\odot}$ year$^{-1}$ before correcting for dust, and
they have starburst spectral energy distributions (Figure~8).  
These systems account for up to 50\% of all
the star formation occurring between $1 < z < 2$, where a large fraction of
the stellar mass in galaxies formed (Dickinson et al. 2003).   The
effective radii of these galaxies varies from 1.5 - 10 kpc, and for these
and other
reasons a fraction of them are likely disks in formation (Conselice et al. 
2004; Elmegreen et al. 2004b; Figure~8).

\begin{figure}[ht]
\centerline{\includegraphics[width=4.5in]{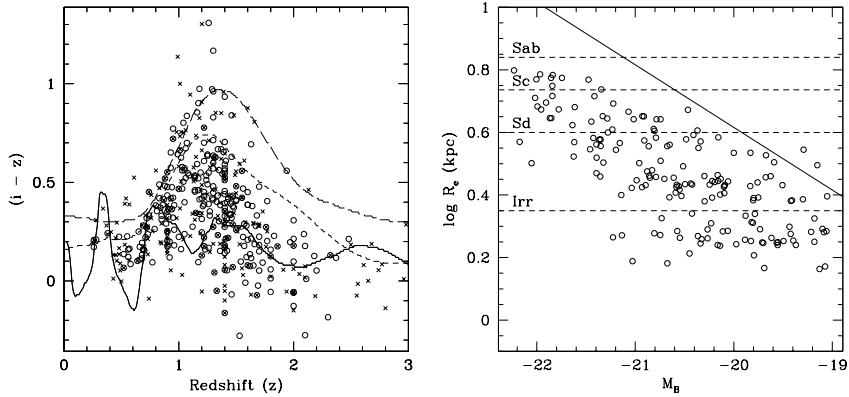}}
\caption{\protect\inx{} Left panel: The distribution of $(i-z)$ colors for 
LDOs (circles) as a function of 
redshift with two Coleman, Wu and Weedman spectral energy distributions and a 
Kinney et al. starburst model plotted (see Conselice et al. 2004a).  These are 
from bluest to reddest - starburst (solid line), Scd (dashed), Sbc (long 
dashed).  Right panel:  Absolute
 magnitude-effective radius relationship for LDOs. The solid
line is the canonical Freeman disk relationship at $z \sim 0$.  The dashed
horizontal lines show the effective radii of different nearby galaxy types.}
\end{figure}

\subsection{Galaxy Mergers}

A major, but still largely under-studied, aspect of galaxy formation is the 
role
of galaxy mergers.   The merging of galaxies
to form larger systems is the cornerstone of the idea behind the modern
galaxy formation model, dark matter, and cosmology (e.g., Cole et al. 2000).
Constraining this process observationally
is just now being done.  The first problem is identifying,
confidently, which galaxies at high redshift are undergoing a major merger 
(defined as a merger mass ratio 3:1 or lower).
There are a few methods for finding galaxies which are merging, or which
soon will. The traditional method for identifying mergers is to identify
galaxies in kinematic
pairs that are close ($< 20$ kpc), and at nearly the same radial velocity
(with $< 500$ km s$^{-1}$).  Identifying pairs at high redshift in this manner
is difficult due to the inability to determine radial velocities for
complete samples, and it is thus largely applicable only at redshifts $z < 1$.

A promising method explored in Conselice (2003) is identifying galaxy mergers 
which are in progress, that is systems that have already merged and are 
undergoing dynamical relaxation.
One way to identify these systems is through their chaotic kinematic
structures revealed through integral field spectroscopy or velocity
curves (e.g., Erb et al. 2004). Alternatively,
and more commonly, is to use the stellar 
structures of galaxies. In \S
2 the methods and reasoning behind using the asymmetry index to find
mergers are explained, and the results of these techniques applied to 
high redshift galaxies are described in \S 3.4.    

Since the modern paradigm for forming galaxies implicitly assumes 
massive galaxies form by merging, it is important to test this idea.  
The agreement between Cold Dark Matter based models and the
data, shown in Figure~6, is good at high redshift, but fails by a significant
amount to reproduce the merger fractions for bright galaxies at lower
redshifts.  This is likely because massive galaxies are forming earlier
than low mass galaxies, which may or may not be an effect of environment - 
massive galaxies are also more likely to be found in dense areas
(Dressler 1980).  We can examine this in more detail to determine how
the stellar masses of high redshift galaxies are built up through mergers and
the star formation induced by this merging.  The amount of stellar mass 
added to
a galaxy during an observed major merger, assuming a mass ratio of 1:1, can be 
calculated from the star forming properties of galaxies 
and the mass accretion rate from merging
(e.g., Papovich et al. 2001; Figure~6).  Based on the merger rates calculated
in Conselice et al. (2003a), a typical Lyman-break galaxy at $z \sim 3$
will undergo $\sim 5$ major mergers, with accompanying star formation,
before $z \sim 1$ and is unlikely to have a major merger at $z < 1$.  The
mass added by each of these mergers, plus the likely amount of new stars
produced in star formation, is enough to create a  $> $ M$_{*}$ (Cole
et al. 2001) galaxy by
$z = 1$.  This also suggest that the galaxy structure-redshift
relationship can be described as a cooling of the galaxy population from
an era of rapid mergers that has been steadily declining since $z \sim 3$.

Minor mergers are harder to constrain, yet are likely a major method for
adding material to normal galaxies at $z < 1$ (e.g., Patton et al. 2002;
Bundy et al. 2004).  This method, or secular
evolution, are the most likely possibilities for driving evolution
in the galaxy population at $z < 1$, when up to 50\% of all
stellar mass formed.  

\subsection{High Redshift Luminous Infrared Galaxies}

A significant galaxy population at high redshift is the luminous infrared 
galaxies
detected by redshifted mid-infrared emission from dust grains at observed
wavelengths 10 - 1000 $\mu$m. The existence of these sources is suggested
by the large far-IR and sub-millimeter backgrounds (Puget et al. 1996).
These luminous infrared sources are very
common at $z > 2$, over an order of magnitude as common as their 
local counterparts  (Chapman et al. 2003a).    With the launch of the
{\it Spitzer Space Telescope}
the study of these galaxies is very much in progress, although important
features of these galaxies are already known.  Luminous infrared
galaxies in the nearby universe are found to be heated by a mixture of
AGN activity and star formation (Sanders \& Mirabel 1996), although
the most luminous sources with L $> 10^{12}$ L$_{\odot}$ are nearly all
starburst-induced major mergers.  The relative contributions at high
redshift is also a mixture of these two types, although preliminary
{\it Spitzer}
observations show that the spectral energy distributions of high
redshift luminous infrared galaxies are largely composed of star forming
systems (Frayer et al. 2004).

The rest-frame UV (or observed optical) properties of sub-mm sources observed
with the Hubble Space Telescope reveal that these galaxies
would be included in magnitude limited optical studies (such as
the HDF) and that their structures are very peculiar in appearance 
(e.g., Chapman et al. 2003b). There are two pieces of evidence that 
suggests these morphologically peculiar sub-mm/radio selected luminous
infrared galaxies at $z > 2$ are undergoing major galaxy mergers. The
first is the very large line widths found in CO mapping (Genzel
et al. 2004). Galaxies with similar large velocity widths in the nearby
universe are undergoing major mergers (Conselice et al. 2000b). 
The other evidence is that the morphologies
of sub-mm sources are generally peculiar (e.g., Chapman et al. 2003b) and
are quantitatively consistent with undergoing major mergers (Conselice
et al. 2003b).  The merger fraction for sub-mm detected galaxies
is in fact higher than for the most massive Lyman-break galaxies (Conselice et
al. 2003b).  This suggests that the most massive galaxies at high redshift,
which are probably these sub-mm sources (Tecza et al. 2004), 
are forming by mergers.

\section{Relationship to the Dark Universe}

The structures of galaxies, and the evolution of structures, potentially 
relates directly with the existence
of dark matter, dark energy and black holes.  I only briefly
discuss this here as much of this is work for the future.
One example is that galaxy mergers might not occur as commonly
without the dynamical friction produced by dark 
matter halos (e.g., Sellwood 2004). Detailed modeling of this however
has not yet been done.   As discussed, black
holes are directly traceable with the concentration of galaxy light (\S 2). 
Dark energy is also likely imprinting its effects, and is
perhaps the fundamental cause of the morphology-redshift relationship.

The relationship between the velocity dispersion of spheroids and the
mass of their central black holes (Gebhardt et al. 2000; Ferrarese \& Merritt
2000) is a fundamental
property of spheroids.  This relationship, which is effectively between
the scale of a spheroid system and its central black hole, is also
projected in the concentration index-black hole mass relationship.  This
relationship appears to hold to some degree up to $z \sim 1.2$ based on the 
correlation
between galaxies with X-ray emission and the CAS concentration index 
(Grogin et al. 2003). X-ray sources up to $z \sim 1.2$ are found in
galaxies with the highest light concentrations, suggesting that
the bulge/central black hole relationship is in place by these redshifts.
Understanding this relationship at higher redshift, as well as how 
dark matter condenses and evolves with galaxy structure, are topics for
future investigations with 20-30 meter ground based telescopes and
the James Webb Space Telescope.

\section{Acknowledgments}

It is a pleasure to acknowledge my collaborators and colleagues, especially
Mark Dickinson, Richard Ellis, Kevin Bundy, and Casey Papovich for
helping shape my evolving understanding of this material.  Thanks also to
Colin Borys and Kevin Bundy for comments on this manuscript.  
I also thank
David Block and the organizing committees for inviting me to present
this contribution and their patience in receiving this review.

\end{document}